\documentclass[final,5p,times,twocolumn,sort&compress]{elsarticle}
\usepackage{amssymb}
\usepackage{color}
\usepackage[dvipsnames]{xcolor}
\colorlet{orange}{orange!120!}
\usepackage{graphicx}% Include figure files
\usepackage{dcolumn}% Align table columns on decimal point
\usepackage{bm}% bold math
\usepackage{amsmath} %Split math lines Verificar se há problemas
\usepackage{amstext}
\usepackage{verbatim}
\usepackage[title]{appendix}
\usepackage[colorlinks,citecolor=blue,linkcolor=blue,anchorcolor=blue,filecolor=blue,urlcolor=blue]{
hyperref}
\usepackage{tabularx}
\usepackage{dcolumn} 
\usepackage{ulem} 
\usepackage{longtable}
\usepackage{soul}
\newcolumntype{d}[1]{D{.}{.}{#1}} 
\definecolor{americanrose}{rgb}{1.0, 0.01, 0.24}
\begin{document}
\begin{frontmatter}
\title{\textbf{Density dependent quark mass model revisited:\\ Thermodynamic consistency, stability windows and stellar properties}}

\author{Bet\^ania C. Backes $^1$}
\author{Eduardo Hafemann $^2$}
\author{Isabella Marzola $^1$}
\author{D\'ebora P. Menezes $^1$}
%%%
\address{
\mbox{$^1$ Departamento de F\'isica, Universidade Federal de Santa
  Catarina, Florian\'opolis, SC, CP 476, CEP 88.040-900, Brazil}\\
\mbox{$^2$ Departamento de Engenharia Qu\'imica e de Engenharia de Alimentos, 
Universidade Federal de Santa Catarina, Florian\'opolis, SC, CP 476, CEP 88.040-900, Brazil}\\}

\begin{abstract} 
In this work a density dependent quark model is
revisited, its thermodynamic consistency checked and the stability window for absolutely stable quark matter obtained. The hypotheses of both pure quark matter with equal quark chemical potentials and stellar matter subject to chemical stability and charge neutrality are investigated. 
The parameters that appear in the density dependent mass and satisfy the Bodmer-Witten conjecture are then used to compute the masses and radii of strange stars. We show that the obtained values are compatible with the recently observed massive stars. 

\end{abstract}
\end{frontmatter}

\section{Introduction}

The solution of quantum chromodynamics (QCD) would give us the key to the understanding of all the physics which depends on the strong nuclear interaction. Unfortunately, this solution is not in sight for the immediate future, what makes the use of relativistic models an important ingredient for the time being.  

To describe the hadronic phase present in the QCD phase diagram, many relativistic mean field models (RMF) are available in the literature. A few years ago, 263 of these models were analysed in \cite{Dutra} and less than forty of them were shown to satisfy nuclear bulk properties completely. Among these models, some of the most successful ones are density dependent hadronic models, whose density dependence is introduced on the meson-baryon couplings as in \cite{DDHD,Typel}. In a RMF approach, a rearrangement term is necessary to account for this dependence when equations of motion and the equation of state (EOS) are calculated so that the thermodynamic consistency is achieved \cite{ours_old}. Another type of dependence is found in a model proposed in 1991 by Brown and Rho (BR) \cite{BR}, in which an in-medium scaling law for the masses and coupling constants for effective chiral Lagrangians was proposed. Within the BR model, both meson masses and couplings are density dependent and a rearrangement term that takes this feature into account is also necessary for the model to satisfy thermodynamic consistency, as shown in \cite{Sidney2006}.

As far as the quark phase is concerned, the most commonly used model is the MIT bag model \cite{mit}, whose name is an acronym for Massachusetts Institute of Technology, where it was conceived. It is a very simple model, where the quarks are described by the Fermi gas EOS and confined through a bag constant, which is introduced to support the gas pressure. Other reference quark model is the Nambu-Jona-Lasinio model \cite{njl}, more robust and capable of describing the chiral symmetry breaking, a property present in QCD. As in hadronic models, quark models that exhibit density dependence, as the QMDD models \cite{qmdd,chakra91} can also be found in the literature. Although the original proposition was revisited many times \cite{chakra93,Peng2000}, these models lack thermodynamic consistency, as shown in \cite{James2013,James_Veronica}.

All terrestrial experiments show that hadrons are the ground state of the strong interaction, but Bodmer and Witten proposed that under specific circumstances, as the ones existing in the core of neutron stars, strange quark matter (SQM) may be the real ground state \cite{Bodmer,Witten}. To test his hypothesis, a stability window is generally searched with the parameters of the models and defined in such a way that a two- flavor quark matter (2QM) must be unstable (i.e., its energy per baryon has to be larger than 930 MeV, which is the iron binding energy) and SQM (three-flavor quark matter) must be stable, i.e., its energy per baryon must be lower than 930 MeV. Taking this conjecture into account, the quark models just cited were tested and the corresponding stability windows found as a function of their free parameters in \cite{James2013} (among many others in the literature). The  Nambu-Jona-Lasinio (NJL) model, which does not have any free parameters, cannot be used in the description of absolutely stable SQM as shown in \cite{seminal,James_Veronica}. Hence, if one believes that a quark star is possible, from the quark models just discussed, only the MIT bag model can be used to describe it. It is important however to stress that the NJL model can still be used in the description of the  quark core inside hybrid stars  \cite{kauan2019,DC2003}. 

In the last decade, very massive neutron stars were confirmed with masses around 2 $M_\odot$, namely J1614-2230 \cite{Demorest} and J0348+0432 \cite{Antoniadis} and possibly reaching 2.84 $M_\odot$ (J0740+6620) \cite{Cromartie}. If one observes Fig.5 in~\cite{James2013}, it is seen that these massive stars cannot be attained with the MIT bag model. 

These considerations made us look for another model that could, at the same time, satisfy thermodynamic consistency and the Bodmer-Witten conjecture and check if massive stars could be obtained. The model that satisfies both conditions is the density dependent mass quark model developed in \cite{Peng2000} and improved in \cite{Xia2014}. In the present work we show that this model does not suffer from the problems present in the QMDD model and can successfully be used to describe massive quark stars.

It is worth stressing that a recent model-independent analysis based on the sound velocity in hadronic and quark matter demonstrated that the existence of quark cores inside massive stars should be considered the standard pattern and not an exotic alternative \cite{nature_2020}. This investigation points to the necessity of a reliable quark matter description. 

The paper is organized as follows: in section \ref{section:formalism}, we summarize the model used to describe SQM, along with the equilibrium conditions necessary to apply the EOS to the study of pure quark matter and stellar matter, necessary to describe compact stars. In section \ref{section:results}, we show the stability conditions for SQM regarding the Bodmer-Witten hypothesis \cite{Bodmer,Witten}, together with our results for the stability windows of both pure quark matter and stellar matter. We then apply the stable parameters to the compact star picture, presenting the maximum masses obtained from the model and thus comparing the values obtained with observational data \cite{Demorest,Antoniadis,Cromartie}. In section \ref{section:final-remarks}, the conclusions are drawn, comparing our results with previous works. In the Appendix \ref{appendix:something}, a way of solving the equilibrium equations with low computational cost is presented.

\section{Formalism at zero temperature}
\label{section:formalism}

In this section, we present the equations derived in Ref. \cite{Xia2014} to describe SQM at zero temperature. In the following model, the masses of the three lightest quarks scale with the baryon number density, mimicking the strong interaction between quarks. We start by considering SQM as a system composed of up ($u$), down ($d$) and strange ($s$) quarks, with the masses given by:

\begin{equation}
    m_{i} = m_{i0} + m_I \equiv m_{i0} + \frac{D}{n_b^{1/3}} + Cn_{b}^{1/3},
    \label{masses}
\end{equation}
where $m_{i0}$ ($i$ = $u$, $d$, $s$) is the current mass of the $i$ quark and $m_I$ is the density dependent quantity that includes the quark interaction effects introduced through the adjustable parameters $C$ and $D$. The 
baryonic density $n_b$ is given by

\begin{equation}
    n_{b} = \frac{1}{3} \sum_i n_i ,
    \label{baryon-number-density}
\end{equation}
where we sum the three quarks densities. At zero temperature, the EOS is obtained once we define the energy density,  

\begin{equation}
    \varepsilon = \Omega_0 - \sum_i \mu_i^* \frac{\partial\Omega_0}{\partial\mu_i^*},
    \label{energy-density}
\end{equation}
and the pressure,

\begin{equation}
    P = -\Omega_0 + \sum_{i,j} \frac{\partial\Omega_0}{\partial m_j} n_i \frac{\partial m_j}{\partial n_i},
    \label{pressure}
\end{equation}
where $\mu_i^*$ is the $i$ quark effective chemical potential and $\Omega_0$ refers to the free unpaired particle contribution given by \cite{Peng2000}:

\begin{equation}
    \Omega_0 = -\sum_i \frac{g_i}{24 \pi^2} \left[ \mu_i^*\nu_i \left( \nu_i^2 - \frac{3}{2}m_i^2 \right) + \frac{3}{2} m_i^4 \ln \frac{\mu_i^* + \nu_i}{m_i} \right],
    \label{thermodynamic-potential-free-system}
\end{equation}
with $g_i$ being the degeneracy factor 6 (3 (color) x 2 (spin)) and $\nu_i$ the fermi momentum of quark $i$, which reads:

\begin{equation}
\nu_i = \sqrt{\mu_i^{*2} - m_i^2},
   \label{eq:fermimom}
\end{equation}
and,

\begin{equation}
    n_i = \frac{g_i}{6\pi^2}(\mu_i^{*2} - m_i^2)^{3/2} = \frac{g_i\nu_i^3}{6\pi^2}.
    \label{particle-density}
\end{equation}

The relation between the chemical potentials and their effective counterparts is:

\begin{equation}
    \mu_i = \mu_i^* + \frac{1}{3}\frac{\partial m_I}{\partial n_b} \frac{\partial \Omega_0}{\partial m_I} \equiv \mu_i^* - \mu_I,
    \label{real-effective-chemical-potentials}
\end{equation}
and the pressure can be rewritten in a more convenient way: 
\begin{equation}
P=-\Omega_{0}+n_{\mathrm{b}} \frac{\partial m_{\mathrm{I}}}{\partial n_{\mathrm{b}}} \frac{\partial \Omega_{0}}{\partial m_{\mathrm{I}}}.
\label{pressure2}
\end{equation}

Given the basic formalism, we proceed to investigate its applications. 
 
\subsection{Pure quark matter}

Although in the literature SQM is usually analysed in the presence of electrons, necessary to ensure chemical equilibrium, this condition is not suitable to the description of stable nuclear matter. Therefore, we choose to apply the formalism to the situation of {\it symmetric matter}, when quarks have equal chemical potentials

\begin{equation}
    \mu_u = \mu_d = \mu_s.
    \label{symmetric-matter}
\end{equation}

Since Eq. (\ref{real-effective-chemical-potentials}) expresses that the real and effective chemical potentials differ only by $\mu_I$, that is the same for every quark flavor, Eq. (\ref{symmetric-matter}) can be rewritten in terms of the effective chemical potentials:

\begin{equation}
    \mu_u^* = \mu_d^* = \mu_s^*.
\end{equation}

One should bear in mind that this definition of symmetric matter is not the same as the usually considered in nuclear matter, where the 
protons and neutrons have equal densities and equal chemical potentials since their bare masses are considered the same. In the  case we analyse here, quarks $u$ and $d$ have similar masses and in the literature they are sometimes considered to be the same, but the strange quark is much more massive. Hence, in this case, the densities of the three quarks are not the same.

\subsection{Stellar matter}

When describing compact stars, we consider SQM to be composed of up ($u$), down ($d$), strange ($s$) quarks and electrons ($e$). Due to the presence of leptons, weak interactions take place, such as

\begin{equation}
    d, s \leftrightarrow u + e + \bar\nu_e.
    \label{decay}
\end{equation}

During the cooling process, the compact star loses a lot of its initial energy by emitting neutrinos, which is known as the Urca process. Restricting our study to cold compact stars, after neutrinos have left the system, the chemical potentials must satisfy $\beta$-equilibrium conditions:

\begin{equation}
    \mu_u + \mu_e = \mu_d = \mu_s.
    \label{chemical-equilibrium}
\end{equation}

Similarly as done for pure quark matter, Eq. (\ref{chemical-equilibrium}) can also be written in terms of the effective chemical potentials:

\begin{equation}
    \mu_u^* + \mu_e = \mu_d^* = \mu_s^*.
    \label{effective-chemical-equilibrium}
\end{equation}

We can see that the effective and true chemicals potentials  of electrons are the same between Eqs.  (\ref{chemical-equilibrium}) and (\ref{effective-chemical-equilibrium}), 
this is because the electrons do not participate in strong interactions, so their corresponding mass is constant.

Compact stars are also considered electrically neutral objects. Accordingly, the charge neutrality condition reads:

\begin{equation}
    \frac{2}{3}n_u - \frac{1}{3}n_d - \frac{1}{3}n_s - n_e = 0.
    \label{charge-neutrality}
\end{equation}

In regards to the EOS, the electrons contribution to the energy density and pressure are considered to be the same as the ones of free fermions. 

In the present formalism, the particle
number densities $n_i$ in Eq. (\ref{charge-neutrality}) and Eq. (\ref{baryon-number-density}) are connected by Eq. (\ref{eq:fermimom}) to the effective chemical potentials $\mu_i^*$, and those are connected with the real chemical potentials
$\mu_i$ by Eq. (\ref{real-effective-chemical-potentials}).
Therefore, Eqs. (\ref{baryon-number-density}), (\ref{chemical-equilibrium}), (\ref{charge-neutrality}), are four equations about the four chemical potentials. Consequently, to find each $n_i$ $(i=u,d,s,e)$ at a given baryon number density $n_b$, it's necessary to solve a nonlinear system with four equations and four unknown variables. It's possible to rewrite this formalism to solve only a nonlinear equation, as shown in Appendix \ref{appendix:something}.\\

\section{Results and discussion}
\label{section:results}

Since QCD is not completely solved yet, it cannot be used to verify the stability of strange matter. Hence, we rely on effective models to describe SQM, evaluating the stability according to the Bodmer-Witten hypothesis \cite{Bodmer,Witten}. From now on, the model above is used to study strange matter, investigating the possibility of stability of SQM against nuclear matter, and after that, we apply the formalism to the description of strange stars.

In order to achieve our goals, the first step is to assume a stability criterion and determine under which parameters strange matter is stable. Therefore, for a large set of $C$ and $D$ parameters, we evaluate the minimum energy per baryon ($\frac{E}{A}$) and classify it into different categories: If it is lower than 930 MeV

\begin{equation}
    \left(\frac{\varepsilon}{n_B} = \frac{E}{A}\right)_{\text{stable SQM}} \leq \text{930 MeV},
    \label{stable}
\end{equation}
that is equivalent to the binding energy of $^{56}$Fe, strange matter is considered to be stable.  If it is higher than 930 MeV, but lower than 939 MeV

\begin{equation}
    \text{930 MeV} <  \left(\frac{\varepsilon}{n_B} = \frac{E}{A}\right)_{\text{metastable SQM}} \leq \text{939 MeV},
    \label{metastable}
\end{equation}
that is equivalent to the mass of nucleons, it is considered to be metastable. If it is higher than 939 MeV, SQM is considered to be unstable

\begin{equation}
    \left(\frac{\varepsilon}{n_B} = \frac{E}{A}\right)_{\text{unstable SQM}} > \text{939 MeV}.
    \label{unstable}
\end{equation}

Simultaneously, the two parameters of the model should be consistent with the fact that two-flavor quark matter (2QM) must be unstable

\begin{equation}
    \left(\frac{\varepsilon}{n_B} = \frac{E}{A}\right)_{\text{unstable 2QM}} > \text{930 MeV},
    \label{2qmstable}
\end{equation}
since matter containing deconfined $u$ and $d$ quarks is found neither in nature nor in terrestrial experiments.  

In face of the model presented, the stability conditions of pure quark matter and stellar matter, and the stability definitions, the energy per baryon is calculated numerically for both strange quark matter and two-flavor quark matter, and then both their minima are evaluated according to the criteria established above.
In what follows, we always consider 
$m_{u0}=$ 5 MeV, $m_{d0}=$ 10 MeV, but the strange quark mass can vary. 

\begin{figure}[h!]
\centering
\includegraphics[scale=0.5]{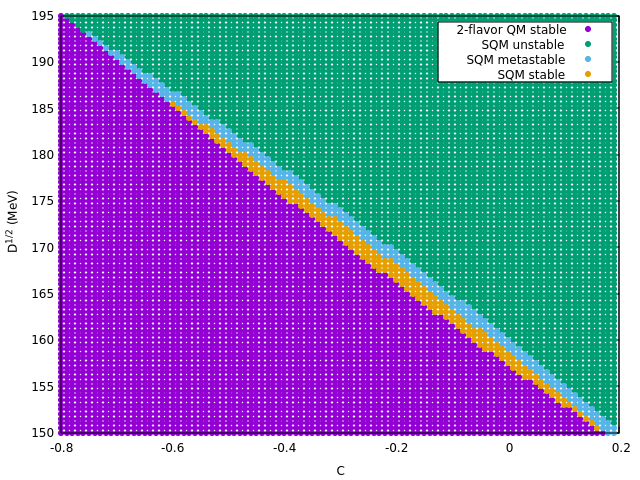}
\caption{(Color online) The stability window for pure quark matter, considering $m_{s0}$ = 100 MeV.}
\label{stability-window-pure-quark-matter-100MeV}
\end{figure}

\begin{figure}[h!]
\centering
\includegraphics[scale=0.5]{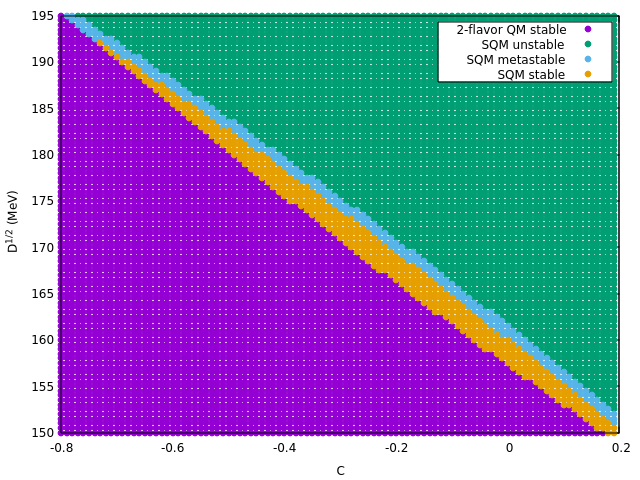}
\caption{(Color online) The stability window for pure quark matter, considering $m_{s0}$ = 80 MeV.}
\label{stability-window-pure-quark-matter-80MeV}
\end{figure}

The stability window for pure quark matter is shown in Figs. \ref{stability-window-pure-quark-matter-100MeV} and \ref{stability-window-pure-quark-matter-80MeV}, considering $m_{s0}$ = 100 MeV and 80 MeV respectively. The lower region is forbidden, corresponding to the sets of parameters on which 2-flavor QM would be stable. SQM is absolutely stable within the yellow and metastable within the blue region. By comparing both results, it is noticeable that Fig. \ref{stability-window-pure-quark-matter-80MeV}, with a smaller value of $m_{s0}$, exhibits a wider range of the parameters that fit stability criteria.

\begin{figure}[h!]
\centering
\includegraphics[scale=0.5]{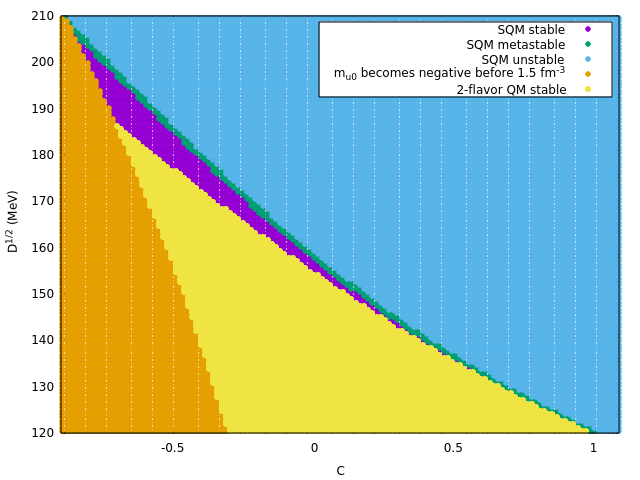}
\caption{(Color online) The stability window for stellar matter with $m_{s0}=$ 100 MeV.}
\label{stability-window-stellar-matter-100MeV}
\end{figure}

\begin{figure}[h!]
\centering
\includegraphics[scale=0.5]{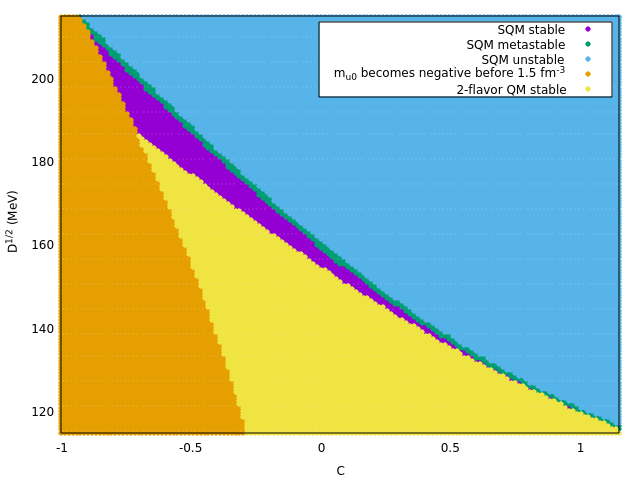}
\caption{(Color online) The stability window for stellar matter with $m_{s0}=$ 80 MeV.}
\label{stability-window-stellar-matter-80MeV}
\end{figure}

For stellar matter, the stability window is presented in Figs. \ref{stability-window-stellar-matter-100MeV} and \ref{stability-window-stellar-matter-80MeV}, also considering $m_{s0}$ = 100 MeV and 80 MeV, respectively. Again, the lower region is forbidden, corresponding to the sets of parameters on which 2-flavor QM would be stable. SQM is absolutely stable within the purple and metastable within the green region. Once more, the stability window obtained is bigger for smaller values of $m_{s0}$. When comparing the results obtained for stellar matter with the ones from pure quark matter, it is noticeable that the the majority of $C$ and $D$ values that satisfy eq. (\ref{stable}) for pure quark matter also satisfy it for the case of stellar matter, with a smaller part satisfying eq. \eqref{metastable} for the latter case.

By analyzing eq. (\ref{masses}), one can notice that for negative values of $C$, the quark masses could become negative at high densities. A negative mass has no physical meaning, resulting in a regime where the model is not valid. Considering it is of interest to apply the model to the study of compact stars, which are objects with very high densities, it is useful to separate the area on which the quark masses become negative at low densities. For this purpose, the region on which $m_{u0}$ becomes negative on densities lower than 1.5 fm$^{-3}$ (around ten times the nuclear saturation density) is detached from the other results shown in Figs. \ref{stability-window-stellar-matter-100MeV} and \ref{stability-window-stellar-matter-80MeV}. 

\begin{figure}[h!]
\centering
\includegraphics[scale=0.5]{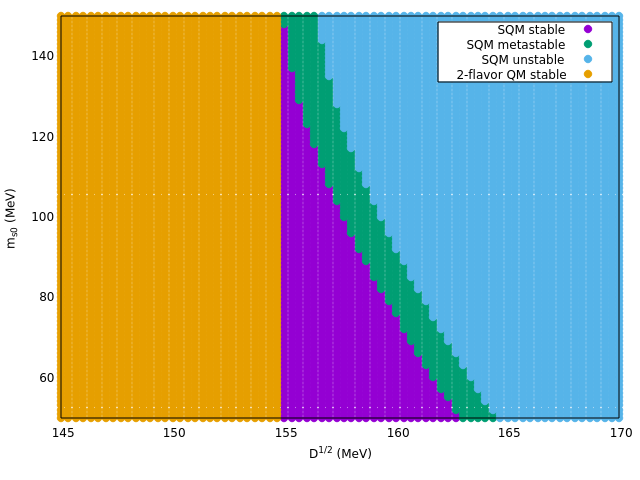}
\caption{(Color online) The stability window for stellar matter. We fix the $C$ parameter as zero in order to observe how the stability window changes with $m_{s0}$.}
\label{stability-window-stellar-matter-ms0}
\end{figure}

Besides the $C$ and $D$ parameters, the quark current masses also have an impact on the EOS. Therefore, it is of interest to map how the stability window changes with the values of the current masses, in particular with $m_{s0}$. In order to do that, we set $C=0$ and numerically calculate the energy per baryon for stellar matter within a wide range of $m_{s0}$ and $D$, which is shown in Fig. \ref{stability-window-stellar-matter-ms0}. It can be seen that smaller values of $m_{s0}$ result in a wider stability range, which is also observed for other values of $C$.

\begin{figure}[h!]
\centering
\includegraphics[scale=0.5]{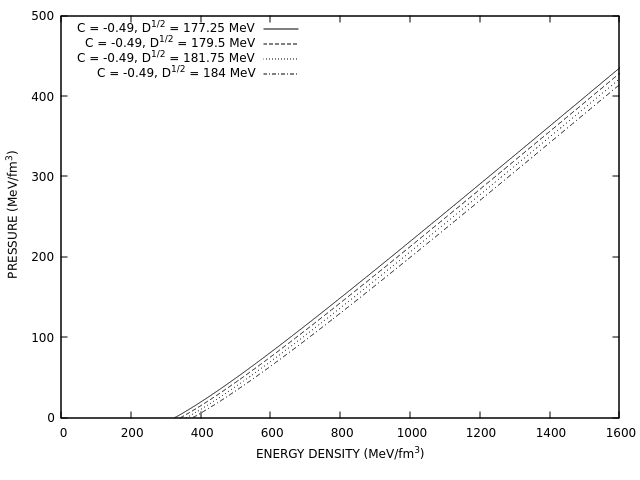}
\caption{The stellar matter equation of state for four different $D$ parameter choices within the stability window, setting $C = -0.49$.}
\label{eos-stellar-matter-c-fixed}
\end{figure}

\begin{figure}[h!]
\centering
\includegraphics[scale=0.5]{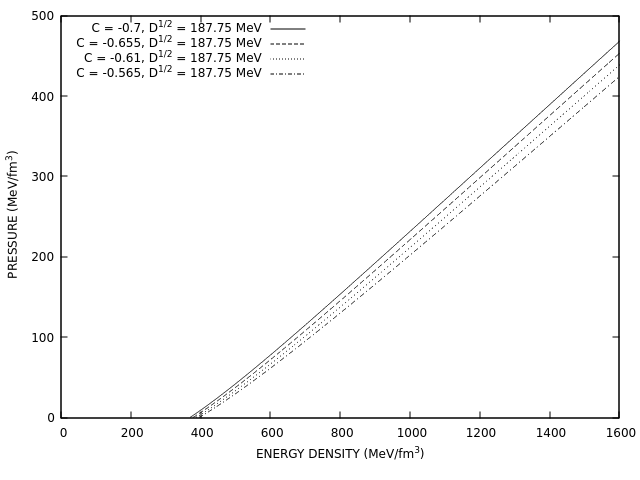}
\caption{The stellar matter equation of state for five different $C$ parameter choices within the stability window, setting $\sqrt{D} = 187.75$ MeV.}
\label{eos-stellar-matter-d-fixed}
\end{figure}

After we obtain the stability windows, it becomes possible to use the stable $C$ and $D$ parameters in order to investigate the EOS more deeply. In Figs. \ref{eos-stellar-matter-c-fixed} and \ref{eos-stellar-matter-d-fixed}, we plot the EOS obtained while maintaining fixed, respectively, the $C$ and $D$ parameters. All the chosen sets of parameters are within the stable region of Fig. \ref{stability-window-stellar-matter-100MeV}, corresponding to stellar matter with $m_{s0} =$ 100 MeV. It is observed that the stiffness of the EOS increases with decreasing values of $C$ and $D$. In addition to that, one can notice that the $D$ parameter has a considerable effect in the zero pressure point, and the parameter $C$ has little influence in this aspect.

\begin{figure}[h!]
\centering
\includegraphics[scale=0.5]{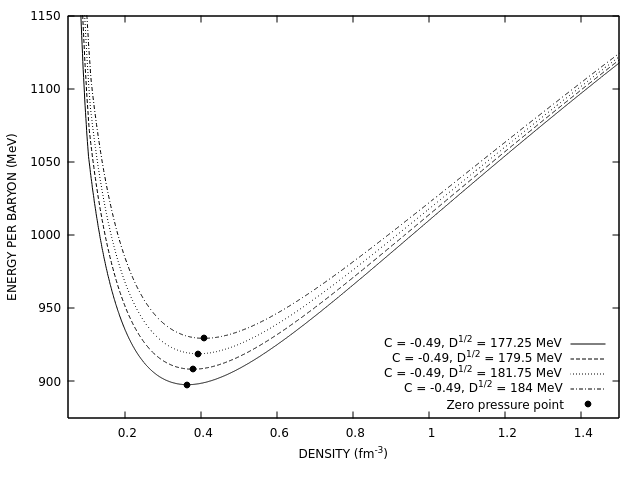}
\caption{The stellar matter energy per baryon as a function of the baryon number density for four different $D$ parameter choices within the stability window, setting $C = -0.49$.}
\label{energy-per-baryon-c-fixed}
\end{figure}

\begin{figure}[h!]
\centering
\includegraphics[scale=0.5]{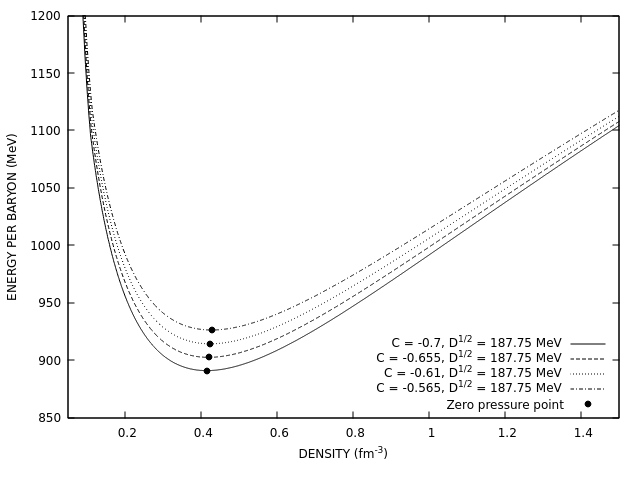}
\caption{The stellar matter energy per baryon as a function of the baryon number density for five different $C$ parameter choices within the stability window, setting $\sqrt{D} = 187.75$ MeV.}
\label{energy-per-baryon-d-fixed}
\end{figure}

A major advantage of the model is its thermodynamic consistency. In order to check if the model satisfies thermodynamic consistency, the energy per baryon as a function of the baryon number density is shown in Figs. \ref{energy-per-baryon-c-fixed} and \ref{energy-per-baryon-d-fixed} for the same sets of parameters used in the previous Figures. One can explicitly see that the binding energy is lower than 930 MeV, satisfying eq. (\ref{stable}), and that
the zero pressure point occurs at exactly the density of the minimum energy per baryon, which is the criterion for thermodynamic consistency, as explained in the Appendix of Ref. \cite{Xia2014} and extensively discussed in \cite{James_Veronica}. Again, we can see the influence of the parameters: as $D$ increases, the energy per baryon decreases and the minimum energy takes place at lower densities; with decreasing values of $C$, the minimum energy decreases.

The model presented in the previous section is an attempt of describing SQM in microscopic situations. Since our goal is to study strange stars at hydrostatic equilibrium and produce an EOS suitable to be used as the core of hybrid stars, we impose the charge neutrality and $\beta$-equilibrium conditions, given by eqs. (\ref{chemical-equilibrium}) and (\ref{charge-neutrality}), and apply them to a given EOS. Then, we have to numerically solve the Tolman-Oppenheimer-Volkoff equations (TOV) \cite{Tolman, Oppenheimer-Volkoff}. The solution of the TOV equations gives the main properties of compact stars, in particular, the mass-radius relation for a family of strange stars at hydrostatic equilibrium.

For $C$ and $D$ parameters along the edges of the stable region in Figs. \ref{stability-window-stellar-matter-100MeV} and \ref{stability-window-stellar-matter-80MeV}, we obtain the radius of the canonical star, the maximum star masses and their respective radii after integrating the TOV equations. These results are presented, respectively, in Tables \ref{tab-max-masses-100MeV} and \ref{tab-max-masses-80MeV}.

\begin{figure}[h!]
\centering
\includegraphics[scale=0.35]{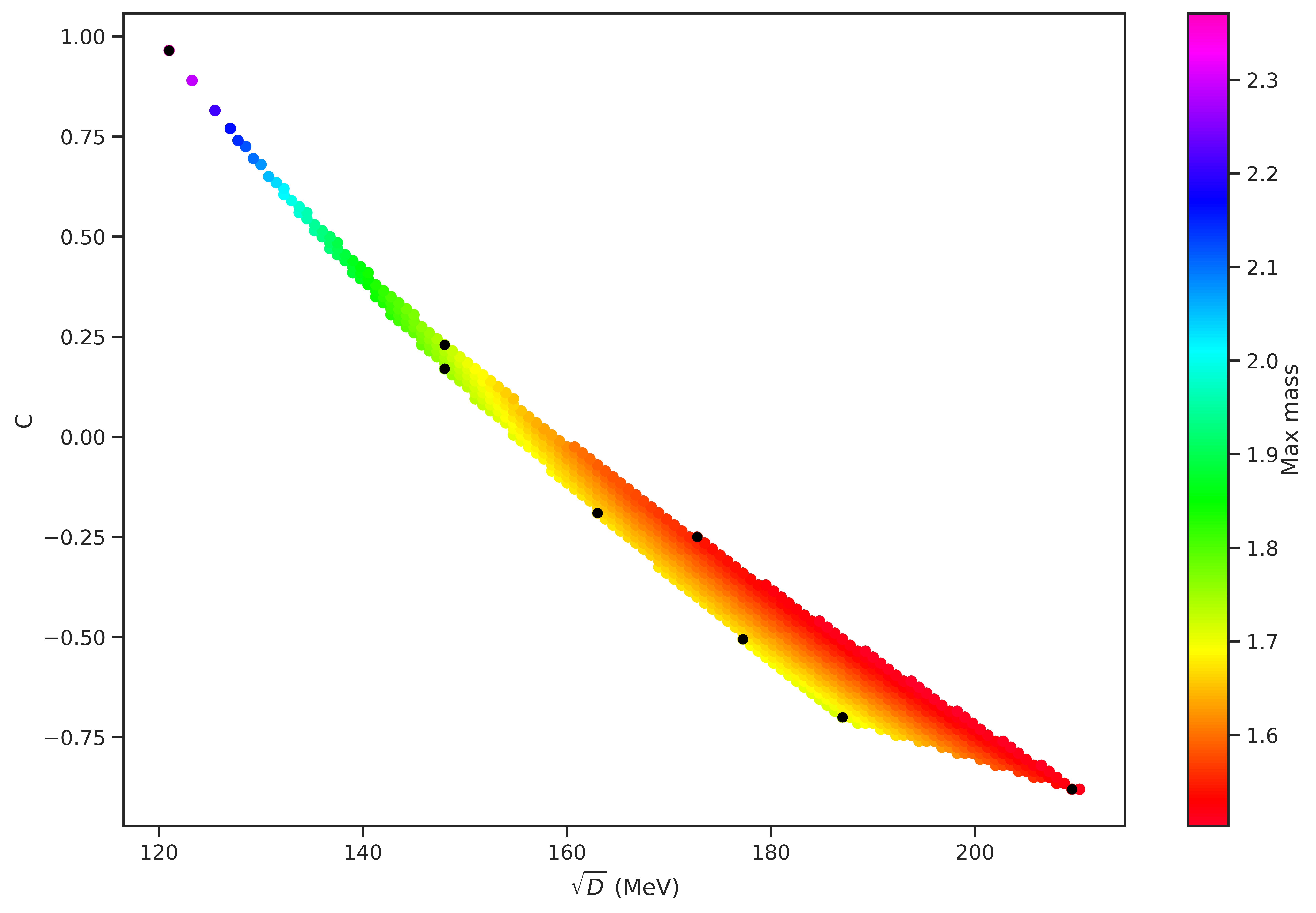}
\caption{(Color online) Maximum stellar masses obtained within the entire stability window, considering $m_{s0}$ = 80 MeV. The black dots correspond to the sets of parameters explicitly shown in Table \ref{tab-max-masses-80MeV}.}
\label{mass-max}
\end{figure}

\begin{figure}[h!]
\centering
\includegraphics[scale=0.35]{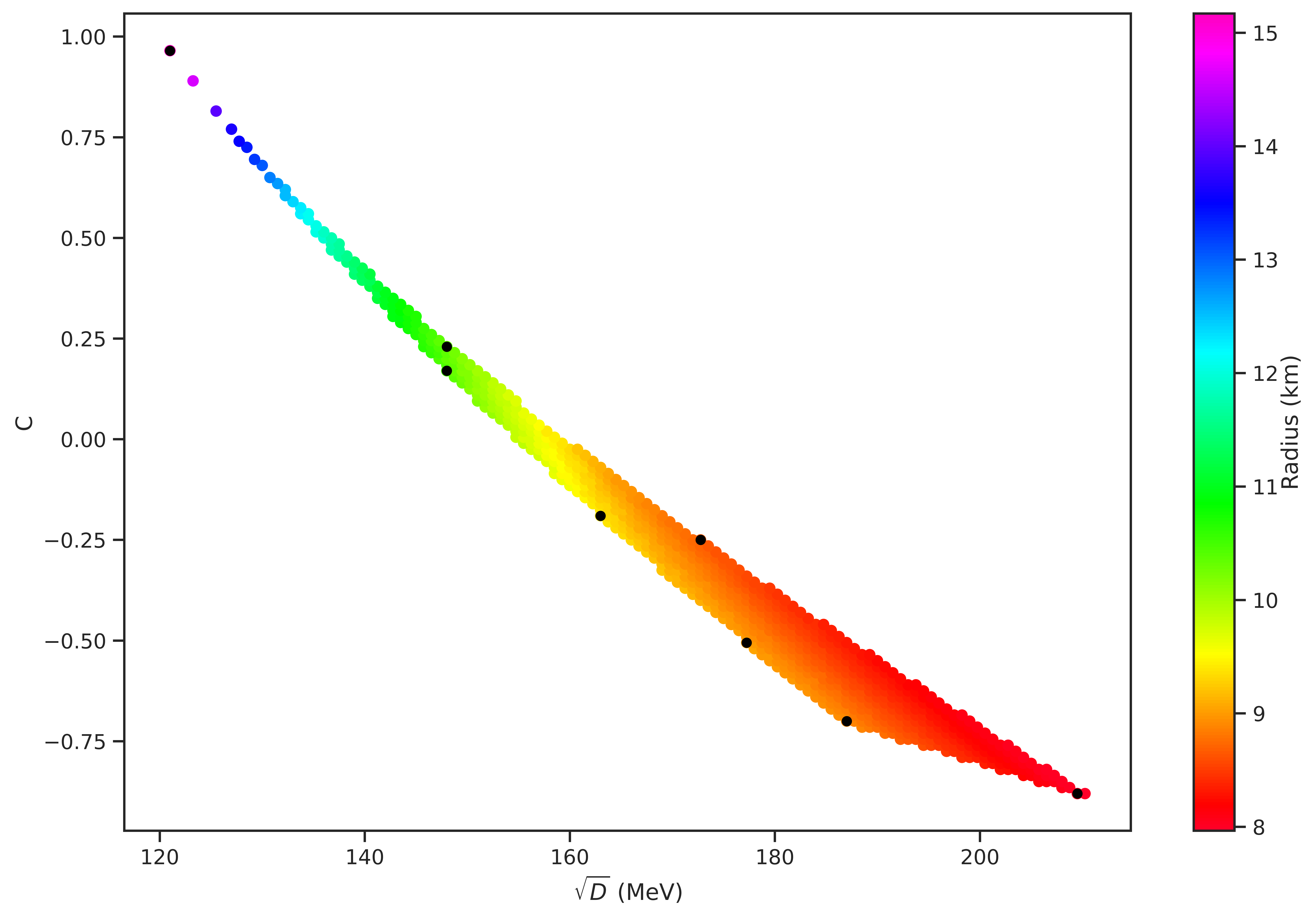}
\caption{(Color online) Radius ofthe star with maximum stellar mass obtained within the entire stability window, considering $m_{s0}$ = 80 MeV. The black dots correspond to the sets of parameters explicitly shown in Table \ref{tab-max-masses-80MeV}.}
\label{radius}
\end{figure}

We graphically show how the maximum masses and their radii vary along the stability window  respectively in Figs. \ref{mass-max} and \ref{radius}. It is observed that the maximum masses increase with decreasing values of the $C$ and $D$ parameters. However, the lower values of $C$ within the stability window are paired with high values of $D$, which makes the higher stellar masses appear in the region with higher values of $C$ and lower $D$. It can be seen in Tables \ref{tab-max-masses-100MeV} and \ref{tab-max-masses-80MeV}, as well as in Figs. \ref{mass-max} and \ref{radius} that there are multiple sets of parameters with very similar stellar masses, but with very different radii. For the majority of the stability window, we find stars with quite small radii, but with maximum masses that are much lower than 2$M_{\odot}$. It is also found that lower values of $m_{s0}$ result in higher maximum masses.

Another important discussion can be made upon the radii of canonical stars, those with 1.4 $M_{\odot}$. A recent study \cite{Capano2020}, based on the binary neutron-star merger GW170817, concludes that the radius of such stars should be between 10.4 and 11.6 km, with a 90\% credibility interval. Such result heavily constrains the EOS of dense matter.

\begin{figure}[h!]
\centering
\includegraphics[scale=0.5]{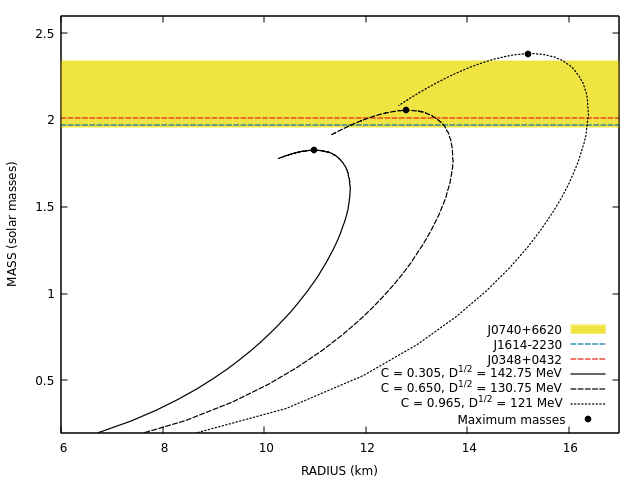}
\caption{(Color online) Mass-radius relation for different strange stars.}
\label{mass-radius}
\end{figure}

The mass-radius diagram for some of the parameter sets is shown in Fig. \ref{mass-radius}, where the maximum masses are marked with full dots. The horizontal lines are the masses of the neutron stars J1614-2230 \cite{Demorest} and J0348+0432 \cite{Antoniadis}, and the wide shaded region covers from the upper to the lower limit of the mass of J0740+6620 \cite{Cromartie} at a 95.4\% credibility interval, all confirmed in the last decade. The set of parameters that gives the highest maximum stellar mass is $C$ = 0.965 and $\sqrt{D}$ = 121 MeV, paired with $ms_{0}$ = 80 MeV, that gives a maximum mass of 2.37 $M_{\odot}$ with a radius of 15.17 km. However, for this set of parameters, the radius of the canonical star ($R_{1.4}$) is 15.53 km, much higher than the upper limit suggested by \cite{Capano2020}. The sets of parameters that satisfy the constraint of $R_{1.4}$ presented in \cite{Capano2020} could not describe stellar masses as high as the ones given by \cite{Demorest,Antoniadis,Cromartie}. From the parameters with $R_{1.4}$ lower than 11.9 km, the highest maximum masses are close to 1.8 $M_{\odot}$, with one of the best set of parameters being $C$ = 0.305, $\sqrt{D}$ = 142.75 MeV, that gives a maximum mass of 1.83 $M_{\odot}$ with $R_{1.4}$ of 11.57 km.

\begin{table}[]
    \centering
    \begin{tabular}{c|c|c|c|c}
    \hline\hline
      $C$ & $\sqrt{D}$ (MeV) & $M_{max}$ ($M_{\odot}$) & $R$ (km) & $R_{1.4}$ (km)  \\ \hline\hline
-0.850 & 205.75 & 1.52 & 8.03 & 8.14 \\ \hline
-0.400 & 178.75 & 1.53 & 8.51 & 8.81 \\ \hline
0.005 & 157.00 & 1.63 & 9.57 & 10.11 \\ \hline
-0.340 & 169.75 & 1.63 & 9.07 & 9.51 \\ \hline
-0.715 & 188.50 & 1.66 & 8.74 & 8.98 \\ \hline
0.080 & 151.75 & 1.70 & 10.02 & 10.63 \\ \hline
0.560 & 133.75 & 1.98 & 12.46 & 13.07 \\ \hline\hline
    \end{tabular}
    
    \caption{The radius of the 1.4 $M_{\odot}$ star ($R_{1.4}$), along with the maximum mass ($M_{max}$) and the respective radius ($R$) for strange stars obtained by integrating the TOV equations with the EOS using $C$ and $D$ parameters along the edges of the stable region of Fig. \ref{stability-window-stellar-matter-100MeV}, with $m_{s0}$ = 100 MeV.
    }
    \label{tab-max-masses-100MeV}
\end{table}

\begin{table}[]
    \centering
    \begin{tabular}{c|c|c|c|c}
    \hline\hline
     $C$ & $\sqrt{D}$ (MeV) & $M_{max}$ ($M_{\odot}$) & $R$ (km) & $R_{1.4}$ (km) \\ \hline\hline

-0.880 &  209.50 & 1.52 & 8.01 & 8.11 \\ \hline
-0.700 & 187.00 & 1.72 & 8.94 & 9.18\\ \hline
-0.505 & 177.25 & 1.69 & 9.03 & 9.39\\ \hline
-0.505 & 187.00 & 1.51 & 8.31 & 8.53\\ \hline
-0.250 & 172.75 & 1.54 & 8.68 & 9.06 \\ \hline
-0.190 & 163.00 & 1.68 & 9.42 & 9.90\\ \hline
0.170 &	148.00 & 1.76 &	10.45 & 11.00\\ \hline
0.230 &	148.00 & 1.73 &	10.35 & 10.92\\ \hline
0.965 &	121.00 & 2.37 &	15.17 & 15.53\\ \hline\hline
    \end{tabular}
    
    \caption{The radius of the 1.4 $M_{\odot}$ star ($R_{1.4}$), along with the maximum mass ($M_{max}$) and the respective radius ($R$) for strange stars obtained by integrating the TOV equations with the EOS using $C$ and $D$ parameters along the edges of the stable region of Fig. \ref{stability-window-stellar-matter-80MeV}, with $m_{s0}$ = 80 MeV.}
    \label{tab-max-masses-80MeV}
\end{table}

\section{Final remarks}
\label{section:final-remarks}

In the present work, we applied the thermodynamically consistent density dependent quark mass model developed in \cite{Xia2014} to the study of pure quark matter, where electrons are absent and the quarks chemical potentials are equal. Similarly to the results presented in \cite{James2013}, we found that the stability window for this case is fairly similar to the one of stellar matter, with charge neutrality and $\beta$-equilibrium, with the stable region being larger for the latter case.

The stability window found for stellar matter is very similar to the one shown in \cite{Xia2014}. In disagreement with the exposed in \cite{Xia2014} is the observation that decreasing values of $C$ imply in stiffer equations of state and, therefore, larger maximum stellar masses, which is counterbalanced with the fact that smaller values of $C$ are only present within the stability window when paired with higher values of $D$.

Regarding the application of the model to the study of strange stars, we notice that the thermodynamically consistent model is capable of describing compact objects more massive than the ones obtained from the MIT Bag Model, but less massive than the ones obtained from the density dependent model that lacks thermodynamic consistency \cite{James2013}. However, the model is capable of describing compact stars with small radii as well as stars with maximum masses up to 2.37 $M_{\odot}$. It is also observed that the same set of parameters cannot simultaneously satisfy both stellar masses as high as \cite{Demorest,Antoniadis,Cromartie} and $R_{1.4}$ as low as the one suggested by \cite{Capano2020}. The equations of state that produce stars with high maximum masses are coupled with $R_{1.4}$ higher than 11.9 km, and the parameters that satisfy this constraint do not produce masses as high as 2 $M_{\odot}$.

The inclusion of temperature in the density dependent mass quark model discussed in the present work is in progress, so that it can be  applied to the description of protoquark stars and also to the investigation of the QCD phase transition as done in \cite{kauan2017}.

\section{Acknowledgments}
This work is a part of the project INCT-FNA Proc. No. 464898/2014-5, partially supported by 
Conselho Nacional de Desenvolvimento Cient\'ifico e Tecnol\'ogico (CNPq) under grant No. 301155/2017-8 (D.P.M.).
B.C.B and I.M. are supported by Coordena\c cão de Aperfei\c comanto de Pessoal de N\'ivel Superior (CAPES) with M.Sc. scholarships and E.H. thanks CNPq for a research initiation scholarship.

\begin{appendices}
\section{Solving the equilibrium conditions equations with a very low computational cost}
\label{appendix:something}

In order to obtain numerical values for the energy density and pressure, one should first determine the densities ($n_i$) and chemical potentials ($\mu_i$) of the particles involved, $i$ being $u$, $d$ and $s$ for pure quark matter or $u$, $d$, $s$ and $e$ for stellar matter. To determine those quantities, the equilibrium conditions should be taken into account, leading into solving a nonlinear equation system. However, when calculating the equation of state multiple times, as when evaluating the EOS for multiple sets of parameters in order to build the stability window of the model, this process can be very time consuming. Nevertheless, it is possible to rewrite the stability conditions, leading into solving only one nonlinear equation, which gives the calculations a lower computational cost, as we expose hereafter for the case of stellar matter.

\subsection{2-Flavor quark matter}

To obtain the stability window of the model, it is also necessary to evaluate the 2-flavor quark matter case, so that points on which 2QM would be stable can be excluded from the stability window. In order to do so, one needs to obtain the particles densities and chemical potentials from the equilibrium equations, as we expose below. We start by rewriting the $\beta$-equilibrium condition in the absence of strange quarks

$$\mu_{u}^{*}+\mu_{e}=\mu_{d}^{*}, $$
as well as the charge neutrality condition

$$\frac{2}{3} n_{u}-\frac{1}{3} n_{d}-n_{e}=0.$$

The relation between the quark effective chemical potentials and densities is the same as before

$$\mu_{i}^*=\sqrt{\left(\frac{6 \pi^{2} n_{i}}{g_{i}}\right)^{2 / 3}+m_i^{2}},$$
where $i$ = $u$, $d$ and the degeneracy factor $g_{i}$ corresponds to 6 (3 (color) x 2 (spin)). For the electrons, the relation is the same, but regarding the real chemical potential

$$\mu_{e}=\sqrt{\left(\frac{6 \pi^{2} n_{e}}{g_{e}}\right)^{2 / 3}+m_e^{2}},$$
with the degeneracy factor $g_e$ being equivalent to 2 (spin).

Lastly, the baryon number conservation reads

\begin{equation}
\frac{1}{3}\left(n_{u}+n_{d}\right)=n_{b}.
\label{baryon-number-2f}
\end{equation}

Rewriting the baryon number conservation in order to isolate $n_{u}$ and replacing the expression into eq. (\ref{baryon-number-2f}), we obtain

\begin{equation}
n_{e}=2n_{b}-n_{d}.
\label{ne-2f}
\end{equation}

We then rewrite the $\beta$-equilibrium condition in terms of the particles densities
$$\sqrt{\left(\pi^{2} n_u\right)^{2 / 3}+m_u^{2}}+\sqrt{\left(3 \pi^{2} n_e\right)^{2 / 3}+m_{e}}=\sqrt{\left(\pi^{2} n_d\right)^{2 / 3}+m_d^{2}},$$
then, it is possible to isolate $n_e$ and rewrite $n_u = 3n_b - n_d$ from eq. (\ref{baryon-number-2f}), resulting in

\begin{multline}
n_{e} = \frac{1}{3 \pi^{2}}
  \left\{ \left[\sqrt{\left(\pi^{2} n_{d}\right)^{2/3}+m_{d}^{2}}-  \right. \right.\\
  \left. \left.
  \sqrt{\left[\pi^{2}\left(3 n_{b}-n_{d}\right)\right]^{2/3}+m_{u}^{2}}\right]^{2}-m_{e}^{2}
\right\}.  
\label{2nd-ne-2f}
\end{multline}

With eqs. \eqref{ne-2f} and \eqref{2nd-ne-2f}, we obtain a nonlinear equation of $n_d(n_b,m_u,m_d,m_e)$ that can be solved numerically in order to find $n_d$. All the other particles densities can then be obtained from the equilibrium conditions, leading to a faster way of obtaining the EOS numerically.
\\

\subsection{Strange quark matter}

The equilibrium conditions applied to the description of the stellar strange quark matter are the same as the ones discussed in Section \ref{section:formalism}. We start by recalling the relation between the particle densities and chemical potentials:

$$n_i=\frac{g_{i}}{6 \pi^{2}}\left(\mu_{i}^{* 2}-m_i^{2}\right)^{3 / 2},
$$

$$n_e=\frac{g_{e}}{6 \pi^{2}}\left(\mu_{e}^{2}-m_{e}^{2}\right)^{3 / 2}.
$$

Rewriting the right side of the $\beta$-equilibrium condition, given by eq. \eqref{effective-chemical-equilibrium}, we have

$${\mu_{d}^*}=\mu_{s}^*.$$

One can now substitute the chemical potentials, using the relation between $\mu_i^*$ and $n_i$, and then isolate $n_d$, obtaining

\begin{equation}
n_{d}=\left[\left(\frac{6 \pi^{2}}{g_{s}} n_{s}\right)^{2/3} +m_{s}^{2}-m_{d}^{2}\right]^{3 / 2}\left(\frac{g_{d}}{6 \pi^{2}}\right).
\label{mid_nd}
\end{equation}

The baryon number conservation, given by eq. \eqref{baryon-number-density}, can be written as
$$n_u=3n_b-n_d-n_s.$$

Replacing $n_d$ by the expression in eq. \eqref{mid_nd} yields
\begin{equation}
n_u =3 n_{b}-\left[\left(\frac{6 \pi^2 n_{s}}{g_{s}}\right)^{2 / 3}+m_{s}^{2}-m_{d}^{2}\right]^{3 / 2}\left(\frac{g_{d}}{6 \pi^{2}}\right)-n_{s}.
\label{mid_nu}
\end{equation}

By considering the charge neutrality condition
$$\frac{2}{3}
n_u-\frac{1}{3}n_d-\frac{1}{3}n_s=n_e,$$
one can then use eq. \eqref{mid_nu} to isolate $n_e$

\begin{equation}
n_{e}=2 n_{b}-\left[\left(\frac{6 \pi^{2}}{g_{s}} n_{s}\right)^{2 / 3}+m_{s}^{2}-m_{d}^{2}\right]^{3 / 2}\left(\frac{g_{d}}{6 \pi^{2}}\right) -n_s.
\label{mid_ne}
\end{equation}

Lastly, recalling the left side from the $\beta$-equilibrium condition in eq. \eqref{effective-chemical-equilibrium}
$${\mu_{u}^*}+\mu_{e}={\mu_{s}^*},$$    
one can rewrite the relation above in terms of the particles densities, which follows

$$\sqrt{\left(\frac{6 \pi^{2} n_u}{g_u}\right)^{2 / 3}+m_ u^{2}}-\sqrt{\left(\frac{6\pi^{2} n_s}{g_s}\right)^{2 / 3}+m_{s}^{2}}=-\sqrt{\left(\frac{6 \pi^{2} n_{e}}{g_{e}}\right)^{2 / 3}+m_{e}^{2}}.$$

For simplicity, we define
$$a_i= \frac{6 \pi^2}{g_i}.$$

Finally, isolating $n_e$, one must find
\begin{multline}
\left\{ \left\{\left[ a_u^{2 / 3}\left(3 n_{b}-\frac{1}{a_d} \left[\left(a_s n_{s}\right)^{2 / 3}+m_{s}^{2}-m_d^{2}\right]^{3 / 2}-n_{s}\right)^{2 / 3}+ m_u^{2}\right]^{1 / 2} -
\right. \right. \\
\left. \left.
\left[\left(a_s n_{s} \right)^{2 / 3}+m_{s}^{2}\right]^{1 / 2}\right\}^{2}-m_e^2 \right\}^{3/2} \frac{1}{a_e} =n_{e}.
\label{final_ne}
\end{multline}

With eqs. \eqref{mid_ne} and \eqref{final_ne}, we have $n_s(n_b, m_u,m_d,m_s,m_e)$. Therefore, at a given baryon number density, since all the quark masses depend only on the density and the free parameters $C$ and $D$, it is only necessary solve a nonlinear equation instead of a nonlinear system in order to find $n_s$. To obtain the other particle densities, eqs. \eqref{mid_nd}, \eqref{mid_nu}, \eqref{mid_ne} can be solved using the value obtained for $n_s$.

The execution time for the numerical calculation of the EOS that uses this method is twice as fast as the nonlinear system. Therefore, the method we propose results in a less expensive computational cost while calculating the EOS multiple times.

\end{appendices}


\begin{thebibliography}{99}

\bibitem{Dutra} M. Dutra, O. Louren\c co, S.S. Avancini, B.v. Carlson, 
A. Delfino, D.P. Menezes, C. Providencia, S. Typel and J.R. Stone,
Phys. Rev. C 90, 055203 (2014).

\bibitem{DDHD} C. Fuchs, H. Lenske and H.H. Wolter, Phys. Rev. C 52 (1995) 3043.

\bibitem{Typel} S. Typel S and H.H. Wolter, Nucl. Phys. A 656 (1999) 331.

\bibitem{ours_old} S.S. Avancini, M.E. Bracco, M. Chiapparini and D.P. Menezes, J. Phys. G: Nuclear and Particle Physics 30 (2004) 27.

\bibitem{BR} G. E. Brown and M. Rho, Phys. Rev. Lett. 66, 2720 (1991).

\bibitem{Sidney2006} S.S. Avancini and D.P. Menezes, Phys. Rev. C 74, 015201 (2006).

\bibitem{mit} A. Chodos, R.L. Jaffe, K. Johnson, C.B. Thorne and V.F. Weisskopf, Phys. Rev. D, 9 (1974) 3471; E. Farhi and R.L. Jaffe, Phys. Rev. D, 30 (1984) 2379.

\bibitem{njl}  Y. Nambu, G. Jona-Lasinio, Phys. Rev. 122, 345 (1961);
Y.Nambu, G.Jona-Lasinio,Phys.Rev.124,246(1961).

\bibitem{qmdd} G.N. Fowler, S. Raha, R.M. Weiner, Z.Phys.C9, 271(1981)

\bibitem{chakra91}
S. Chakrabarty, Phys. Rev. D, 43 (1991) 627.

\bibitem{chakra93}
S. Chakrabarty, Phys. Rev. D 48 (1993) 1409. 

\bibitem{Peng2000} 
G.X. Peng, H.C. Chiang, B.S. Zou, P.Z. Ning and S.J. Luo, Phys. Rev. C, 62 (2000) 025801.

\bibitem{James2013}
J.R. Torres and D.P.Menezes, Europhysics Letters, 101 (2013) 42003.

\bibitem{James_Veronica}
V. Dexheimer, J.R. Torres and D.P. Menezes,
Eur. Phys. Jour. C 73:2569 (2013) 

\bibitem{Bodmer} A.R. Bodmer, Phys. Rev. D 4, 1601 (1971).

\bibitem{Witten}  E. Witten, Phys. Rev. D, 30 (1984) 272.

\bibitem{seminal}  D.P. Menezes, M. Benghi Pinto, S.S. Avancini, 
A. P\'erez Martinez and C. Provid\^encia -  Phys. Rev. C 79, 035807 (2009).

\bibitem{kauan2019} C.A. Graeff, M.D. Alloy, K.D. Marquez,
C. Providencia and D.P. Menezes,  JCAP 01 (2019) 024.

\bibitem{DC2003} D.P. Menezes and C. Provid\^encia, Phys. Rev. C 68 (2003) 035804. 

\bibitem{Demorest} P.B. Demorest et al,
Nature, 467, 1081 (2010).

\bibitem{Antoniadis}
J.~Antoniadis et al: Science 340, 1233232 (2013)

\bibitem{Cromartie}
H.~Cromartie et al: Nat. Astr. (2019)

\bibitem{Xia2014} C.J. Xia, G.X. Peng, S.W. Chen, Z.Y. Lu and J.F. Xu,
  Phys. Rev. D 89, 105027 (2014). 
  
\bibitem{nature_2020} E. Annala, T. Gorda, A. Kurkela et al.  Nat. Phys. (2020).
  
\bibitem{Tolman}
R.C. Tolman, Phys. Rev 55, 364 (1939)

\bibitem{Oppenheimer-Volkoff}
J.R. Oppenheimer and G.M. Volkoff, Phys. Rev 55, 374 (1939)

\bibitem{kauan2017} K.D. Marquez and D.P. Menezes, JCAP 12(2017) 028.

\bibitem{Capano2020}
C.D. Capano, I. Tews, S.M. Brown et al, Nat Astron 4, 625–632 (2020)

\end{thebibliography}
\end{document}